\documentclass[a4paper]{article}
\usepackage{float}
\usepackage{multirow}
\usepackage{amssymb}
\usepackage{xcolor}
\usepackage{INTERSPEECH2021}
\usepackage{bm}

\usepackage[colorlinks=true,citecolor=green]{hyperref}
\usepackage{cleveref}

\title{The SJTU System for Short-duration Speaker Verification Challenge 2021}
\name{Bing Han, Zhengyang Chen, Zhikai Zhou, Yanmin Qian\textsuperscript{\dag} \thanks{\textsuperscript{\dag}Yanmin Qian is the corresponding author}}

\address{
MoE Key Lab of Artificial Intelligence, AI Institute \\
X-LANCE Lab, Department of Computer Science and Engineering \\
Shanghai Jiao Tong University, Shanghai, China
}
\email{\{hanbing97, zhengyang.chen, zhikai.zhou, yanminqian\}@sjtu.edu.cn}

\begin{document}
\maketitle
\begin{abstract}
This paper presents the SJTU system for both text-dependent and text-independent tasks in short-duration speaker verification (SdSV) challenge 2021. In this challenge, we explored different strong embedding extractors to extract robust speaker embedding.
For text-independent task, language-dependent adaptive snorm is explored to improve the system performance under the cross-lingual verification condition. For text-dependent task, we mainly focus on the in-domain fine-tuning strategies based on the model pre-trained on large-scale out-of-domain data. In order to improve the distinction between different speakers uttering the same phrase, we proposed several novel phrase-aware fine-tuning strategies and phrase-aware neural PLDA. With such strategies, the system performance is further improved. Finally, we fused the scores of different systems, and our fusion systems achieved 0.0473 in Task1 (rank 3) and 0.0581 in Task2 (rank 8) on the primary evaluation metric. 

\end{abstract}

\section{Introduction}
Speaker verification system has gained great improvement with the development of deep learning. From the phone-channel \cite{sre} to in-the-wild condition \cite{voxceleb_eval}, researchers have proposed different architectures \cite{xvector,resnet,ecapa,dpn}, different losses \cite{aam,angular,ge2e}, and different training strategies \cite{chung2020defence,zhang2017end,adversarial} to improve the system performance under different conditions. However, there are still some challenges unresolved when the speaker verification system is applied in the real world, such as the short duration problem and cross-lingual problem.

In this paper, we introduce the SJTU system submitted to the short-duration speaker verification (SdSV) challenge 2021~\cite{sdsvc2021plan}. The SdSV challenge 2021 includes two tasks. The task1 is the text-dependent task, where the speaker verification system should verify the test speaker identity and speaking phrase at the same time. Task2 is text-independent task, the system should only consider the speaker identity. Particularly, the SdSV challenge introduces a new challenging verification condition, the cross-lingual verification for task2, where one speaker may speak different languages at the enrollment and test stage.

The SdSV 2021 is the second challenge of SdSV series and many competitive systems have been proposed in the last challenge. For text-independent task in the last challenge, Jenthe et al. proposed a new data mining strategy HPM \cite{HPM} and introduced adaptive snorm to improve system's cross language verification robustness. Peng et al. introduced a greedy fusion algorithm \cite{greedyfusion} to further improve the performance of the fusion system. Besides, teams mainly focused on the back-end optimization \cite{netease, but_sdsvc2020} in text-dependent task.

In this challenge, we first explored different well-performed architectures and trained them on all the available data. Then, we focus on the in-domain data fine-tuning strategies to further improve the system performance. To solve the cross-language verification problem in text-independent task, we trained another language identification network to introduce the language information to the adaptive s-norm \cite{snorm} procedure. For text-dependent task, we implemented different methods to increase the distinction between the target trial and different non-target trials. We used an ASR system to classify the speaker phrase during the test stage and filter out the phrase-mismatch (speaker utters a wrong pass-phrase) trials directly. To better distinguish different speakers uttering the same phrase, we proposed several novel phase-aware fine-tuning strategies and phrase-aware neural PLDA. Based on such strategies, the performance of our systems is further improved. 

The rest of the paper is organized as follows: Section 2 introduces the dataset used in this challenge. Section 3 introduces our embedding extractor architectures and the proposed fine-tuning strategies. The experimental results and corresponding analysis are given in Section 4. Finally, we make the conclusion in Section 5.

\section{Datasets}

\subsection{Training Data}

SdSV challenge adapted a fixed training condition where the system should only be trained on the designed set. The main training and evaluation data for SdSV challenge is the DeepMine \cite{deepmine1, deepmine2} dataset which was recorded in realistic environments of Iran. And the collection protocol was designed to incorporate various kinds of noises during the recording. The main language is Persian while the most of the participants also participated in the English partition.

\begin{itemize}
    \item Task 1 in-domain data: A dataset which is designed for building text-dependent speaker verification system. It consists of 101k utterances from 963 different speakers. The content of all utterances is limited to a fixed set including five Persian phrases and five English phrases.
    \item Task 2 in-domain data: A dataset which has no restrictions on utterance content. It contains 125k utterances collected from 588 speakers while some of them have only Persian phrases.
\end{itemize}

In addition to the in-domain training data, other opening datasets allowed to be used in the training process are described as follows.

\textbf{Voxceleb} \cite{voxceleb1,voxceleb2}: Voxceleb 1\&2 contain more than one million utterances from 7245 celebrities, which are collected from videos uploaded to YouTube. 

\textbf{Librispeech} \cite{librispeech}: A dataset which comprises 281k utterances from 2338 speakers. It's sourced from audio books and the majority of speech is US English.

\textbf{Common Voice Farsi} \cite{commonvoice}: The Common Voice corpus is a massively-multilingual collection of transcribed speech. And only Persian part of it is used in this challenge.



\subsection{Evaluation}
The evaluation data for task 1\&2 are both part of the DeepMind in-domain data.

\begin{itemize}
    \item In Task 1, each trial consists of a test segment along with a model identifier which indicates three enrollment utterances and a phrase ID that uttered in the utterances. These trials can be classified into four basic types including TC, TW, IC and IW. The text-dependent speaker verification system should accept the TC trials and reject the other three types as imposture trials.
    \item In Task 2, the enrollment data consists of one to several variable-length utterances from the Persian language while the test utterances might from the different language (English). For this task, systems should accept the trials if enroll and test utterances are both from the same speaker without considering language mismatch.
\end{itemize}

The main metric adopted by SdSV challenge is normalized minimum detection cost function (minDCF), which is defined as a weighted sum of false alarm and miss error probabilities.

\section{Methods}

In this section, we will introduce the embedding extractors, fine-tuning strategies and several post-processing methods used in our system. In our experiment, the embedding extractors are firstly trained on all the available data for both task1 and task2 in a text-independent mode. Then, we fine-tune the pre-trained models using in-domain data. Finally, post-processing methods are used to further improve the system performance.

\subsection{Speaker Embedding Extractors}
\label{ssec:emb_extractor}

To build a robust speaker verification system for SdSV challenge, all datasets including Voxceleb, Common Voice Farsi, Librispeech, and DeepMind in-domain data are combined for training the speaker embedding extractors. In order to reduce the duration mismatch between the training and test data, all the utterances are randomly chunked into segments of 2 seconds during the training stage. The acoustic feature we used is 40 dimensional Fbank with 25 ms frame length and 10 ms shift. To increase the quantity and diversity of the training data, we apply online data augmentation during the training process. The additive noises from MUSAN corpus \cite{musan} and the impulse responses from RIR \cite{rir} are used for augmentation.



In our system, we mainly adopt three different speaker verification architectures, including the ResNet34 \cite{r-vector}, ECAPA-TDNN \cite{ecapa} and DPN68 \cite{dpn, dpn68}.

\textbf{ResNet34}: ResNet has achieved superior performance in speaker verification for its high-efficiency modeling complex data structure. We use the ResNet34 introduced in \cite{r-vector} as our resnet based architecture. In this architecture, input features are processed by the initial convolution layer and 4 residual blocks,  then a following statistic pooling layer aggregates the frame-level features into segment-level representation. Finally, a 256-dimensional fully connected layer transforms it into a fixed vector to represent the speaker.

\textbf{ECAPA-TDNN}: The ECAPA-TDNN \cite{ecapa} has achieved great success in speaker verification system and has been used in the VoxSRC2020 \cite{voxsrc2020} winning system. We set the channel number for ECAPA-TDNN to 1024 in our experiment. Channel attention with and without global context are both applied and we denote the corresponding architectures as Ecapa and Ecapa-Glob respectively.


\textbf{DPN68}: The DPN (Dual Path Network) \cite{dpn} is firstly applied in speaker verification task in \cite{dpn68}, which leverages the advantage of ResNet and DenseNet \cite{densenet} at the same time. Here, we use the DPN68 architecture as one of our embedding extractors in our systems.

Additive angular margin softmax loss (AAM) \cite{aam} is used to optimize all the embedding extractors. The scale parameter and the margin of AAM loss are set to 32 and 0.2 respectively. We train each model for 165 epochs and the learning rate exponentially decreases from 0.1 to 1e-5 during the training process.



\subsection{In-domain Fine-tuning}
In this section, we will introduce our fine-tuning strategies based on the pre-trained model presented in the last section to further improve the system performance on the in-domain evaluation set.



\begin{figure}[h]
  \centering
  \includegraphics[width=0.8\linewidth]{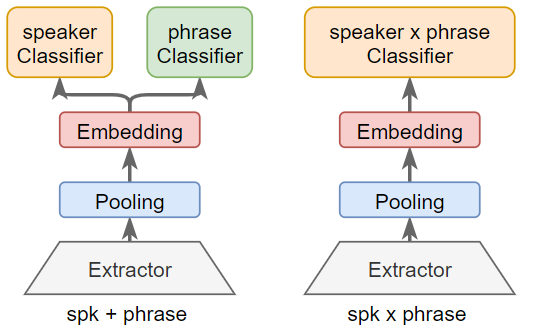}
  \caption{Text-dependent Mode Fine-tuning for Task 1}
  \vspace{-0.4cm}
  \label{fig:finetune_1}
\end{figure}

\subsubsection{Text-Dependent Mode Fine-tuning for Task 1}
\label{sssec:text_dependent_mode}
To encode the phrase information into speaker embedding and further enlarge the distance of different speakers uttering the same phrase, we fine-tune the embedding extractors in text-dependent mode for task1. Strategies are introduced as follows:


\textbf{spk} $+$ \textbf{phrase}: As shown in Figure.~\ref{fig:finetune_1} (left), in the fine-tuning stage, there are two separate heads for speaker and phrase classification and we fine-tune the embedding extractors in a multi-task way.


\textbf{spk} $\times$ \textbf{phrase}: Here, utterances in different phrases spoken by the same speaker are considered as the different classes. As shown in Figure.\ref{fig:finetune_1} (right), there is only one classification head, but both speaker and phrase information are considered.


Since the classification of the phrase requires all the information of a sentence, the inputs are not chunked during the training process and variable-length inputs in the same batch are zero-padded to the same length.

\subsubsection{Text-Independent Mode Fine-tuning for Task1 and Task2}
\label{sssec:text_independent_mode}

\begin{figure}[h]
  \centering
  \includegraphics[width=0.8\linewidth]{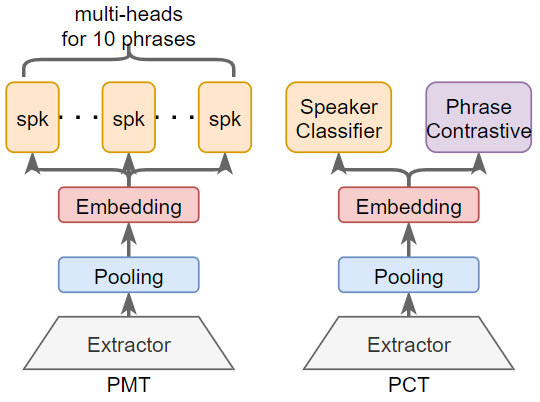}
  \caption{Text-independent Mode Fine-tuning for Task 1}
  \label{fig:finetune}
  \vspace{-0.4cm}
\end{figure}

In our experiment, the phrase mismatch trials (IW and TW) in task1 can be filtered out by ASR system introduced in section \ref{sssec:asr_sys}. Therefore, the model only needs to verify utterances in the same phrase, and task1 can also be regarded as text-independent task. For usual fine-tune methods of task1 and task2, we use AAM softmax to optimize the pre-trained models on in-domain data. 

Especially for task1, to strengthen model's ability to discriminate speakers uttering the same phrase, we proposed two phrase-aware text-independent fine-tuning strategies, including phrase-aware multi-head training (PMT) and phrase-aware contrastive training (PCT).

\textbf{PMT}: For task 1, all the phrases are drawn from a fixed set of ten phrases consisting of five Persian and five English phrases. As shown in Figure \ref{fig:finetune} (left), different speaker classification heads are used for the utterances in different phrases. With such a training strategy, the distance between different speakers within the same phrase can be enlarged.

\textbf{PCT}: As shown in Figure \ref{fig:finetune} (right), in this fine-tuning strategy, we introduce a contrastive learning loss that can be jointly optimized with the AAM softmax loss. In our experiment, generalized end-to-end loss \cite{ge2e} is adopted to calculate the contrastive loss and we sample two utterances for each speaker in the training batch. To improve the distinction between different speakers uttering the same phrase, we constrain that all the utterances in the same batch are from the same phrase.


\subsection{Post Processing}

\subsubsection{Language-dependent AS-Norm}
The most difficulty of task 2 is cross-language trials. To minimize the language mismatch between the different utterances, we introduce the language information to the adaptive symmetric score normalization (AS-Norm), which is defined in Equation \ref{eq:asnorm}. The cohort set for each enroll model, $\mathcal{E}_{e,lan}^{top}$, is decided by the enroll model and the language of test utterance, where the language of cohort set is the same as test utterance. To detect the language of the test utterances, a TDNN based language identification is trained on task 2 in-domain data.

\begin{equation}
\label{eq:asnorm}
    s(e, t) = \frac{s(e,t)-\mu(S_t(\mathcal{E}_t^{top}))}{\sigma(S_t(\mathcal{E}_t^{top}))}+\frac{s(e,t)-\mu(S_e(\mathcal{E}_{e,lan}^{top}))}{\sigma(S_e(\mathcal{E}_{e,lan}^{top}))} \nonumber
\end{equation}
\subsubsection{Phrase-aware Neural PLDA}
\label{sssec:nplda}

The neural PLDA (NPLDA) \cite{nplda} was successfully applied in the NICT system \cite{nict_sdsvc} of SdSV challenge 2020. To enhance the NPLDA for text-dependent task, we constrain the input pair for NPLDA from the same phrase to improve the different speakers' distinction ability within the same phrase. Our NPLDA was initialized by phrase-dependent PLDA model trained on task2 in-domain data and the parameters were set the same as \cite{nplda}, except learning rate = 5e-5 and epochs = 5. 


\subsubsection{ASR System}
\label{sssec:asr_sys}

For text-dependent task1, an ASR system is trained to filter the trials that enroll and test utterances are from different phrases. We adopt the conformer~\cite{conformer} based joint CTC-attention automatic speech recognition model~(ASR) from the ESPnet~\cite{watanabe2018espnet} Librispeech recipe. The ASR system is first trained on the Librispeech dataset and then fine-tuned on the task1 in-domain data. During evaluation, we use the ASR system to recognize the phrases and classify each utterance based on its Levenshtein edit distance with the references. According to the phrase label generated by ASR, we directly filter out IW and TW trials by setting the score to a very low value. It is noted that all the results for task1 provided in the experiments are revised based on this ASR system.


\section{Experiments}

\subsection{Task 1: Text-Dependent Speaker Verification}

\subsubsection{Pre-trained Model}

All the embedding extractors introduced in section \ref{ssec:emb_extractor} are firstly pre-trained on all the available datasets and the corresponding results are listed in Table \ref{tab:feature_model}. From the results, we can see that ResNet34 using cosine similarity measurement obtains the best performance. In comparison with Ecapa-TDNN based models, DPN68 also exhibits a better performance but worse than ResNet34. In addition, the cosine scoring method outperformed the PLDA in most cases, and we will only provide the cosine result in the following sections for analysis.

\begin{table}[H]
  \caption{Results comparison of Pre-trained Models on Task 1}
  \label{tab:feature_model}
  \centering
\begin{tabular}{lcccc}
\toprule
\multirow{2}{*}{Model Type} & \multicolumn{2}{c}{Cosine}       & \multicolumn{2}{c}{PLDA} \\
                            & EER            & minDCF          & EER        & minDCF      \\
\midrule
ResNet34                      & \textbf{2.680} & \textbf{0.0811} & 2.680      & 0.1032      \\
Ecapa                       & 3.093          & 0.0994          & 3.093      & 0.0984      \\
Ecapa-Glob                  & 2.887          & 0.1008          & 3.093      & 0.0943      \\
DPN68                         & 2.680          & 0.0882          & 3.299      & 0.0984      \\
\bottomrule
\end{tabular}
\end{table}

\subsubsection{In-domain Fine-tuning}
\label{sssec:task1_indomain_finetune}
\begin{table}[b]
\vspace{-0.4cm}
  \caption{Text-dependent Mode Fine-tuning of Task 1}
  \label{tab:td_finetune}
  \centering
\begin{tabular}{ccc}
\toprule
Text-dependent Finetune                 & EER            & minDCF         \\
\midrule
-                       & 2.680          & 0.0811          \\ 
speaker $\times$ phrase          & 2.474          & 0.0740           \\ 
speaker $+$ phrase          & \textbf{2.268} & \textbf{0.0718} \\ 
\bottomrule
\end{tabular}
\end{table}

\begin{table*}[htbp]
  \vspace{-0.4cm}
  \caption{Main Results of Task 2. Fine-tune is applied on Task 2 in-domain data with AAM softmax}
  \label{tab:task2}
  \centering
\begin{tabular}{lcccccccc}
\toprule
\multirow{2}{*}{Model} & \multicolumn{2}{c}{ResNet34} & \multicolumn{2}{c}{Ecapa} & \multicolumn{2}{c}{Ecapa-Glob} & \multicolumn{2}{c}{DPN68} \\ 
                       & EER           & minDCF        & EER         & minDCF       & EER            & minDCF         & EER         & minDCF       \\ 
\midrule
Pre-trained                                     & 2.657         & 0.1036        & 2.793       & 0.1252       & 2.929          & 0.1281         & 2.452       & 0.0975       \\ 
{ \ + \ \ Fine-tune}                                    & 2.316         & 0.0933        & 2.793       & 0.1173       & 2.589          & 0.1219         & 2.316       & 0.0832       \\ 
{ \ ++ A-snorm}                                     & \textbf{1.981}         & 0.0872        & 2.520       & 0.0991       & 2.316          & 0.1045         & 2.112       & \textbf{0.0752}       \\ 
\bottomrule
\end{tabular}
\vspace{-0.4cm}
\end{table*}

\textbf{Text-Dependent Mode}: Table \ref{tab:td_finetune} illustrates the comparison of different text-dependent mode fine-tuning strategies introduced in \ref{sssec:text_dependent_mode}. Limited by GPU memory and time consumption, large models with whole utterances as input are difficult to train. Here, experiments are only based on Resnet34. From the result, we can see that fine-tuning will bring improvement to the text-dependent task. Especially, ``speaker + phrase" obtained the best performance among these results.

\textbf{Text-Independent Mode}: Our proposed text-independent mode phrase-aware fine-tuning strategies are investigated in this section and the results are shown in Table \ref{tab:tid_finetune}. It's obvious that all fine-tuning systems perform better than the pre-trained model. In this table, we also list the fine-tuning results with PCT (phrase-aware contrastive training) and PMT (phrase-aware multi-head training) for all models. Compared with the usual fine-tuning method on the in-domain data, PCT and PMT both achieve excellent performance improvement on both EER and minDCF. In addition, ResNet34 with PCT strategy performs the best within all the models.

\begin{table}[h]
  \caption{Text-independent Phrase-aware Fine-tune for Task 1. The usual Fine-tune without PCT and PMT is also done on Task 1 for the comparison where the in-domain data is trained with AAM softmax in text-independent mode.}
  \label{tab:tid_finetune}
  \centering
\begin{tabular}{lccccc}
\toprule
Model                                                                 & Fine-tune & PCT & PMT & EER   & minDCF \\ 
\midrule
\multirow{4}{*}{ResNet34}                                             &          &    &     & 2.680 & 0.0811 \\  
                                                                      & \checkmark        &     &     & 2.680 & 0.0734 \\ 
                                                                      & \checkmark        & \checkmark   &     & \textbf{2.260} & \textbf{0.0642} \\ 
                                                                      & \checkmark        &     & \checkmark   & 2.680 & 0.0708 \\ \hline
\multirow{4}{*}{Ecapa}                                                &          &     &     & 3.093 & 0.0994 \\ 
                                                                      & \checkmark        &     &     & 3.093 & 0.0992 \\ 
                                                                      & \checkmark        & \checkmark   &     & 2.680 & 0.0863 \\ 
                                                                      & \checkmark        &     & \checkmark   & 2.887 & 0.0852 \\ \hline
\multirow{4}{*}{\begin{tabular}[l]{@{}l@{}}Ecapa\\ Glob\end{tabular}} &          &     &     & 2.887 & 0.1008 \\ 
                                                                      & \checkmark        &     &     & 2.680 & 0.0863 \\ 
                                                                      & \checkmark        & \checkmark   &     & 2.680 & 0.0855 \\ 
                                                                      & \checkmark        &     & \checkmark   & 3.093 & 0.0772 \\ \hline
\multirow{4}{*}{DPN68}                                                &          &     &     & 2.680 & 0.0882 \\ 
                                                                      & \checkmark        &     &     & 2.887 & 0.0866 \\ 
                                                                      & \checkmark        & \checkmark   &     & 2.680 & 0.0759 \\ 
                                                                      & \checkmark        &     & \checkmark   & 2.680 & 0.0759 \\ 
\bottomrule
\end{tabular}
\end{table}

\subsubsection{Phrase-aware Neural PLDA}
As mentioned in Table \ref{tab:feature_model}, PLDA cannot provide a satisfactory performance compared to cosine similarity. To further enhance the fusion system performance, we also trained the phrase-aware NPLDA introduced in section \ref{sssec:nplda} to improve the result of PLDA which can be used in the final fusion system. We conduct this investigation based on the models fine-tuned with PCT strategy which obtain the best result in section \ref{sssec:task1_indomain_finetune} and the corresponding NPLDA result can be found in Table \ref{tab:nplda}. Compared with traditional PLDA, NPLDA shows its effectiveness and brings an excellent improvement on minDCF which is comparable with cosine back-end results.

\begin{table}[h]
  \caption{NPLDA Results of Task 1}
  \label{tab:nplda}
  \centering
\begin{tabular}{lcccc}
\toprule
\multirow{2}{*}{Model}           & \multicolumn{2}{c}{PLDA} & \multicolumn{2}{c}{NPLDA} \\
      & EER        & minDCF      & EER         & minDCF      \\
\midrule
ResNet34   & 3.093      & 0.1060      & 2.887       & 0.0734      \\
Ecapa      & 2.887      & 0.1106      & \textbf{2.680}       & 0.0844      \\
Ecapa-Glob & 2.887      & 0.1021      & 2.887       & 0.0738      \\
DPN68      & 3.505      & 0.1009      & 2.887       & \textbf{0.0656}      \\
\bottomrule
\end{tabular}
\vspace{-0.4cm}
\end{table}

\subsection{Task 2: Text-Independent Speaker Verification}

The main results of task2 are listed in the Table \ref{tab:task2}. In the task2, all the embedding extractors are also first pre-trained on all the available datasets and then fine-tuned on the in-domain data. According to the results, we can see that DPN68 achieves the best result within all the models. Besides, compared with the pre-trained models, fine-tuning with in-domain data delivers a significant performance improvement. Besides, we also conduct an investigation of language-dependent asnorm. DPN68 with asnorm outperforms others with the lowest minDCF 0.0752. 

\subsection{Fusion Result}

\begin{table}[h]
\vspace{-0.2cm}
  \caption{Fusion Results on Dev and Eval Set}
  \label{tab:final_result}
  \centering
\begin{tabular}{cccccc}
\toprule
\multirow{2}{*}{Task}                      & \multicolumn{2}{c}{Dev set} & \multicolumn{2}{c}{Eval set} &  \multirow{2}{*}{Rank}   \\ 
                  & EER      & minDCF        & EER         & minDCF        &                   \\ 
\midrule
Task1                 & 2.268           & 0.0493         & 1.44           & 0.0473             & 3                     \\ 
Task2                 & 1.703           & 0.0579         & 1.23           & 0.0581             & 8                     \\ 
\bottomrule
\end{tabular}
\vspace{-0.2cm}
\end{table}

Finally, the scores from all the systems including different models, back-ends and fine-tuning strategies are weighted summed to get the fusion system and we use the development set for fusion weights tuning. The results of fusion systems on the development set and evaluation set are shown in Table \ref{tab:final_result}. From the table, we can see that a fusion system could further improve the performance. Our primary submission is produced by fusion systems and achieved 0.0473 in Task1 (rank 3) and 0.0581 in Task2 (rank 8) on minDCF.

\section{Conclusion}
In this paper, we give a detailed description of our submission to Task 1 \& 2 of SdSV Challenge 2021. Several strong embedding extractors are explored in our experiment. For text-independent task, another language identifier is used to introduce language information to the adaptive snorm. For text-dependent task, an ASR system is used to filter the IW and TW trials. We proposed several phrase-aware fine-tuning and post-processing methods to strengthen model's ability to verify speakers within the same phrases. Based on these strong systems, our final fusion system achieved $3^{rd}$ and $8^{th}$ place in task1 and task2 respectively.

\section{Acknowledgements}

This work was supported by the China NSFC projects (No. 62071288 and No. U1736202). Experiments have been carried out on the PI super- computer at Shanghai Jiao Tong University. And the author would like to thank the support of Houjun Huang and Xu Xiang from AISpeech Ltd. 

\bibliographystyle{IEEEtran}

\bibliography{mybib}


\end{document}